\documentclass[a4paper, 11pt]{article}

\def\b{\begin{eqnarray}}
\def\e{\end{eqnarray}}
\def\n{\noindent}

\def\p{\partial}

\def\f{\varphi}

\def\f{\phi}

\def\2{\frac{1}{A^2 - \phi}}

\def\phj{\phantom{j}}

\begin{document}

\begin{center}

{\huge \textbf{Haunted Kaluza Universe with \\ Four-dimensional Lorentzian Flat, \\ Kerr, and Taub--NUT Slices \\}}
\vspace {10mm}
\noindent
{\large Rossen I. Ivanov$^\dagger$ and \setcounter{footnote}{6}
Emil M. Prodanov$^\ast$}
\vskip1cm
\n
\hskip-.3cm
\begin{tabular}{c}
\hskip-1cm
$\phantom{R^R}^\dagger${\it Department of Pure and Applied Mathematics, Trinity College,} \\ {\it University of Dublin,
Ireland, and} \\
{\it Institute for Nuclear Research and Nuclear Energy, Bulgarian} \\ {\it Academy of Sciences,
72 Tsarigradsko chaussee, Sofia -- 1784, Bulgaria} \\ {\it e-mail: ivanovr@tcd.ie} \\
\\
\hskip-.8cm
$\phantom{R^R}^\ast${\it School of Physical Sciences, Dublin City University, Glasnevin, Ireland} \\
{\it e-mail: prodanov@physics.dcu.ie}
\end{tabular}
\vskip1cm
\end{center}

\vskip2cm

\begin{abstract}
\n
The duality between the original Kaluza's theory and Klein's subsequent modification is duality between slicing and
threading decomposition of the five-dimensional spacetime. The field equations of the original Kaluza's theory lead
to the interpretation of the four-dimensional Lorentzian Kerr and Taub--NUT solutions as resulting from static
electric and magnetic charges and dipoles in the presence of ghost matter and constant dilaton, which models
Newton's constant.

\end{abstract}

\newpage

\section{Introduction}

What is referred today as Kaluza--Klein theory (see \cite{Review} for an extensive collection of papers) is Klein's
modification \cite{Klein} of the original Kaluza's theory \cite{Kaluza}. These two theories are dual \cite{ip1, ip2}
and the duality between them is the duality between threading and slicing decomposition \cite{Boersma} of the
five-dimensional spacetime --- foliation with one-dimensional or one-codimensional surfaces. \\
The field equations \cite{Thiry} of Klein's theory (threading decomposition) express Newton's constant as a dynamical
field (dilaton) and do not allow a constant solution for the dilaton unless an unphysical restriction to the Maxwell
tensor $F_{ij}$ is imposed: namely, $F_{ij} F^{ij} = 0$ (Latin indexes run from 1 to 4, Greek --- from 1 to 5).
In 1983 Gross {\it et al.} \cite{Gross} and Sorkin \cite{Sorkin} found magnetic monopoles in Klein's theory by
considering four-dimensional Euclidean and periodic in time Kerr \cite{Kerr} and Taub--NUT \cite{Taub, NUT} solutions
which were trivially embedded into a vacuum five-dimensional Klein's universe with timelike fifth dimension. The
original Euclidean periodic time was then identified as the fifth dimension and the magnetic vector potentials --- as
former degrees of freedom of the four-dimensional Kerr or Taub--NUT solution. The resulting four-dimensional gravity
has a non-constant dilaton and has lost the original Kerr or Taub--NUT geometry. \\
Unlike Klein's theory, in the original Kaluza's theory (slicing decomposition) the gauge degrees of freedom of the
electromagnetic potentials $A_i$  are transferred to the dilaton \cite{ip1, ip2}. This allows us to consider
four-dimensional spacetimes with constant dilaton (i.e. Newton's constant) and fixed gauge. We show
that a constant dilaton and a vacuum five-dimensional Kaluza universe necessitate a Ricci-flat four-dimensional slice.
In the dual Kaluza's setup, the Kerr or Taub--NUT geometry of the four-dimensional slice is preserved. The field
equations of the original Kaluza's theory lead to the interpretation of the four-dimensional Lorentzian Kerr and
Taub--NUT solutions as resulting from static electric and magnetic charges and dipoles in the presence of ghost matter.

\section{Field Equations of Kaluza's Theory}

\noindent
The five-dimensional Kaluza's metric is:
\b
\label{ka}
G_{\mu \nu} = \left( \begin{array}{ccc|c}
 & & & \\
 & g_{ij}^{\phantom{A^A}} & & A_i  \\
 & &  & \\
\cline{1-4}
& A_i^{\phantom{A^A}} & & \f  \\
\end{array}  \right) \,
\e
The five-dimensional interval in mostly-plus metric is:
\b
\label{inte}
d
\sigma^2 = g_{ij} (dy^i + A^i d s)(dy^j + A^j d s) + \frac{1}{N^2} ds^2 \, ,
\e
where $y^1 \equiv t \, , \, y^5 \equiv s \, , \, g^{ij}$ is the inverse of $g_{ij} \, , \, A^i = g^{ij} A_j$ and
$N^{-2} = \f - A^2$. \\
The slicing lapse function is $N^{-1}$, while the slicing shift vector field is given by $A^i$. \\
If one is to require $g_{ij}$ to be the metric of our four-dimensional world and  $A_i$ --- the electromagnetic
potentials, then $N$ is the dilaton field and can be expressed as \cite{ip1}:
\b
N^2 = \frac{\mbox{det } g}{\mbox{det } G} \, .
\e
The five-dimensional Kaluza metric $G_{\mu \nu}$ is a solution to the five-dimensional vacuum equations
$R_{\mu \nu} = 0$, where $R_{\mu \nu}$ is the five-dimensional Ricci tensor. These equations were written in terms of
the extrinsic curvature $\pi_{ij} = -(N/2) (\nabla_i A_j + \nabla_j A_i - \p_s g_{ij}) $ of the four-dimensional
world as follows \cite{ip1, ip2, Wesson}:
\b
r_{ij} - \frac{1}{2}g_{ij} r & = & N \nabla_i \nabla_j \frac{1}{N} - N (\mathcal{L}_{\mbox{\tiny A}} \pi_{ij}
+ \p_s \pi_{ij}) + (\pi \pi_{ij} - 2 \pi_{ik} \pi^k_j) \nonumber \\
& & \hskip4.99cm + \frac{1}{2} g_{ij} (\pi^2 - \pi_{kl} \pi^{kl}) \, , \\
0 & = & \nabla_i (\pi^i_j - \delta^i_j \pi) \, , \\
\lower2pt \hbox{$\phantom{.}^{\mbox{\tiny \fbox{}}}$} \frac{1}{N} &
= & A^i \nabla_i \pi - \frac{1}{N} \pi_{ij} \pi^{ij} - \p_s \pi \, ,
\e where $r_{ij}$ is the four-dimensional Ricci tensor, $r$ is the
four-dimensional scalar curvature, $\: \nabla_i$ is the
four-dimensional covariant derivative, $\lower1pt
\hbox{$\phantom{.}^{\mbox{\tiny \fbox{}}}$} = g^{ij} \nabla_i
\nabla_j \, ,$
$\:\: \mathcal{L}_{\mbox{\tiny A}}$ is the Lie derivative with respect to $A_i$ and  and  $\pi = \pi^k_k$. \\
These equations can be equivalently written as \cite{ip1, ip2}:
\b
\label{einstein}
& & r_{ij} - \frac{1}{2} g_{ij}r  =  \frac{N^2}{2} T_{ij} \, , \\
\label{maxgen}
& & \nabla_i F^{ij} = - 2 A_i r^{ij} + \frac{2}{N^2} (\pi^{ij} - \pi^k_k g^{ij}) \p_i N \, , \\
\label{div}
& & \nabla_i (A_j \pi^{ij} - \nabla^i \frac{1}{N}) = 0 \, , \\
\e
where $F_{ij} = \p_i A_j - \p_j A_i$ is the Maxwell electromagnetic tensor. \\
The first of these equations, (\ref{einstein}), are the equations of general relativity with matter, the second,
(\ref{maxgen}), are a generalization of Maxwell's equations and the last, (\ref{div}), is the gauge-fixing condition.
The dilaton $N$ is related to Newton's constant $G_N$ via \cite{ip1}:
\b
\frac{N^2}{2} = \kappa = \frac{8 \pi G_N}{c^4} \, .
\e
The energy-momentum tensor $T_{ij}$ appearing in equation (\ref{einstein}) is given by \cite{ip1, ip2}:
\b
T_{ij} = T^{\mbox{\tiny Maxwell}}_{ij} \, + \, \nabla^k \Psi_{ijk} \, +
\, \nabla^k \Theta_{ijk} \, + \, C_{ij} \, + \, D_{ij} \, ,
\e
where:
\b
T^{\mbox{\tiny Maxwell}}_{ij} & = & F_{ik} F^{\phj k}_j - \frac{1}{4} g_{ij} F_{kl} F^{kl} \, , \\ \nonumber \\
\Psi_{ijk} & = & A_k \nabla_j A_i - A_j \nabla_k A_i + A_i F_{jk} \, , \\ \nonumber \\
\Theta_{ijk} & = & \nabla_i (A_k A_j) + g_{ij}(A^l \nabla_k A_l - A_k \nabla_l A^l)  \, ,  \\ \nonumber \\
C_{ij} & = & g_{ij} A^k A^l r_{kl} - 2 A^k A_j r_{ik} - 2 A^k A_i r_{jk} \, , \\ \nonumber \\
D_{ij} & = & \frac{2}{N} \nabla_i \nabla_j \frac{1}{N} - \frac{2}{N^2} \pi^k_k (A_i \p_j N + A_j \p_i N)
\nonumber \\
& & \!\!\!\!\! + \frac{2}{N^2} \Bigl[ - A^k \pi_{ij} + A_i \pi^k_j + A_j \pi^k_i - g_{ij}
    (A^l \pi^k_l - A^k \pi^l_l)\Bigr] \p_k N.
\e
The field equations reveal a very interesting relation between the type of solution of the four-dimensional general
relativity and the dilaton. \\
We first suppose that the dilaton is constant: $N = \mbox{const}$. Let us write the five-dimensional vacuum
metric $G_{\mu \nu}$ in a block-diagonal form:  $G_{\mu \nu} = \mbox{diag }(g'_{ij}, \, N^{-2})$. Having $A_i = 0$
in the field equations, together with $N = \mbox{const}$, results in vanishing of the full energy-momentum tensor and,
therefore in a Ricci-flat four-dimensional relativity ($r_{ij} = 0$). One can re-introduce the electromagnetic
potentials via a five-dimensional coordinate transformation. The only transformation which leaves the five-dimensional
interval (\ref{inte}) invariant is: $y^i \rightarrow y^i + s c^i, \: s \rightarrow s, \, $ where $c^i$ are constants.
Then the physical electromagnetic potentials will be $A_j = g_{jk} c^k$ . Under this transformation the fields in
the five-dimensional interval (\ref{inte}) transform as $g'_{ij} = g_{ij} \, , \: A'^i = A^i + c^i \, , \: N' = N$.
Thus $r_{ij} = 0$ remains unchanged. Therefore constant dilaton and a vacuum five-dimensional Kaluza universe
necessarily result in a Ricci-flat four-dimensional slice.

\section{Flat Four-dimensional Universe}

It is interesting to consider whether the converse is true, namely, if a Ricci-flat four-dimensional slice
($r_{ij} = 0$), embedded in a vacuum five-dimensional Kaluza universe ($R_{\mu \nu} = 0$), results in a constant
dilaton ($N = \mbox{const}$). We will give an example which shows that this is not the case. Consider a flat
four-dimensional slice with $g_{ij} = \eta_{ij} = \mbox{diag}(-1, \, 1, \, 1, \, 1)$. This is clearly a vacuum
solution. Let us now see if the five-dimensional metric $G_{\mu \nu} = \mbox{diag}(-1, \, 1,
\, 1, \, 1, \, N^{-2})$ admits a non-constant solution for $N$. From the field equations we see that when
$r_{ij} = 0$ and $A_i = 0$, then $N$ must be a solution to:
\b
\nabla_i \nabla_j \frac{1}{N} = 0 \, .
\e
For the flat case, the obvious solution is: $N = (a_k y^k + a_5)^{-1}$, where $a_\mu$ are constants. It will be very
interesting to take:
\b
\label{kas} a_2 = a_3 = a_4 = a_5 = 0 \, , \:\: a^1 = \frac{c^2}{4 \sqrt{\pi} t_0} \, .
\e
Then Newton's constant will become $G_N = c^4 N^2/(16 \pi) = (t_0/t)^2$ --- the gravitational attraction falling off
with time from infinity. This can be explained as a purely geometric effect between a vacuum universe embedded into
another vacuum universe. \\
The Kasner metric \cite{Kasner} is:
\b
d \sigma^2 = -dt^2 + \sum_{i=2}^4 \left( \frac{t}{t_0} \right)^{2p_i} (dy^i)^2 + \left(\frac{t}{t_0} \right)^{2p_5}
ds^2 \, ,
\e
where $\sum_{i=2}^5 \: p_i = \sum_{i=2}^5 \: p_i^2 = 1$. \\
Solution (\ref{kas}) corresponds to the special case: $p_2 = p_3 = p_4 = 0 \, , \, p_5 = 1$.

\section{Ghost Energy-Momentum Tensor}

As we are interested in solutions to the vacuum five-dimensional relativity with constant four-dimensional Newton's
constant (dilaton), we will have to consider Ricci-flat solutions to four-dimensional relativity only. For $r_{ij} = 0$
and $N = \mbox{const}$ the field equations reduce to:
\b
\label{e1}
r_{ij} - \frac{1}{2} h_{ij}r & = & \frac{N^2}{2} (T^{\mbox{\tiny Maxwell}}_{ij} + \nabla^k \Psi_{ijk}
+ \nabla^k \Theta_{ijk}) \, = \, 0 \, , \\
\label{e2}
\nabla_i F^{ij} & = & 0 \, , \\
\label{e3}
\nabla_i (A_j \pi^{ij}) & = & 0 \, .
\e
We further have:
\b
\label{eqm1}
\nabla^j T^{\mbox{\tiny Maxwell}}_{ij} = F_{ik} \nabla_j F^{jk} = 0 \, ,
\e
due to (\ref{e2}). Equation (\ref{eqm1}) is the conservation law for the energy and momentum resulting from Maxwell's
equations (\ref{e2}). \\
The tensor $\nabla^k \Theta_{ijk}$ satisfies:
\b
\label{eqm2}
\nabla^j \nabla^k \Theta_{ijk}  = - \frac{2}{N} \nabla_j \nabla_i (A_k \pi^{ik}) = 0 \, ,
\e
in view of the gauge-fixing condition (\ref{e3}). \\
Considering the remaining term, $\nabla^k \Psi_{ijk}$, we see that it satisfies:
\b
\label{eqm3}
\nabla^j \nabla^k \Psi_{ijk} & = & \frac{1}{2} (\nabla^j \nabla^k + \nabla^k \nabla^j) \Psi_{ijk} +
\frac{1}{2} [\nabla^j, \nabla^k] \Psi_{ijk} = 0
\e
in view of the antisymmetry $\Psi_{ijk} = - \Psi_{ikj}$ and $r_{ij} = 0$. Thus the tensor $\nabla^k \Psi_{ijk}$ does
not describe any dynamics. \\
The gauge-fixing condition (\ref{e3}) and equation (\ref{eqm3}) lead to the conservation law:
\b
\nabla^j T^{\mbox{\tiny Ghost}}_{ij} = 0 \, ,
\e
where $T^{\mbox{\tiny Ghost}}_{ij}$ is the Belinfante symmetric energy-momentum tensor of the ghost fields:
\b
T^{\mbox{\tiny Ghost}}_{ij} = \nabla^k (\Psi_{ijk} + \Theta_{ijk}) \, .
\e
For the "haunted" Kaluza's universe, the energy and momentum of the ghost fields compensates completely the energy
and momentum of the Maxwell's fields:
\b
T^{\mbox{\tiny Ghost}}_{ij} + T^{\mbox{\tiny Maxwell}}_{ij} = 0
\e
and, therefore, it is possible to have matter co-existing with ghost matter in a Ricci flat universe.

\section{Four-dimensional Lorentzian slice with Kerr \\ Geometry}

We will generate the five-dimensional solution simply by
starting off with a four-dimensional static Ricci-flat solution, promoting it trivially to five dimensions
(by adding $ds^2$ in the metric) and performing a five-dimensional coordinate transformation:
\b
\label{shift}
t \rightarrow t + \beta s \, ,
\e
where $\beta$ is the inverse of the ``speed of light'' along the fifth, transverse dimension. This coordinate
transformation will not introduce $s$-dependence in the four-dimensional world (as the four-dimensional metric is
static and time appears only with its differential) and as a result the new {\it five-dimensional} "observer" will
``see'' the electromagnetic potentials:
\b
\label{pot}
A_j = \beta g_{tj} \, ,
\e
where $j \in \{ t, \, r, \, \theta, \, \phi \}$. \\
The four-dimensional Kerr metric \cite{Kerr} in Boyer--Lindquist coordinates \cite{Boyer}, trivially promoted to
five dimensions is:
\b
d \sigma^2 & = & - \frac{\Delta}{\rho^2} (dt - a \sin^2 \theta \,\, d \phi)^2
+ \frac{\sin^2 \theta}{\rho^2} [(r^2 + a^2) \, d\phi - a \, dt]^2 \nonumber \\
& & \hskip30pt + \, \frac{\rho^2}{\Delta} \, dr^2 + \rho^2 \, d\theta^2 + ds^2 \, ,
\e
where $\Delta = r^2 - \alpha r + a^2$ and $\rho^2 = r^2 + a^2 \cos^2 \theta$. Here $\alpha$ and $a$ are integration
constants which will be identified further (in the four-dimensional Kerr solution these are the mass and the angular
momentum per unit mass of a black hole). We consider the physically interesting case $\alpha > a$ (black hole
solution). \\
The electromagnetic potentials (\ref{pot}) are:
\b
A_t & = & \beta (-1 + \frac{\alpha r}{\rho^2}) \, , \\
A_r & = & 0 \, , \\
A_\theta & = & 0 \, , \\
A_\phi & = & - \frac{a \alpha \beta r \sin^2 \theta}{\rho^2} \, .
\e
For large $r$, the non-zero components of the vector potential are:
\b
\label{el}
A_t & \sim & \beta (-1 + \frac{\alpha}{r}) \, , \\
\label{mag} A_\phi & \sim & - \, \frac{a \alpha \beta \sin^2
\theta}{r} \, .
\e
Therefore, from (\ref{el}), one can identify the constant $\alpha \beta$ as electric charge:
\b
\label{qkerr}
q = \alpha \beta \, .
\e
Equation (\ref{mag}) describes the field of a magnetic dipole of strength $a \alpha \beta$, located at the
origin \cite{Gross, Sorkin}:
\b
\label{mkerr}
m = a \alpha \beta = a q \, .
\e
Thus one can interpret the Kerr solution as a black hole generated by an electric charge and magnetic dipole (and
not by a rotating massive centre). The potentials (\ref{pot}) satisfy the vacuum Maxwell's equations (\ref{e2}).
The electric charge and the magnetic dipole are located at the singularity $\rho = 0$ . \\
This is the only singularity of the Kerr spacetime and can be better understood in Cartesian coordinates \cite{ch}:
\b
x & = & \sqrt{r^2 + a^2} \, \sin \theta \, \cos \phi \, , \\
y & = & \sqrt{r^2 + a^2} \, \sin \theta \, \sin \phi \, , \\
z & = & r \cos \theta \, . \\
\e
The singularity $\rho = 0$, i.e. $r = 0$ and $\cos \theta = 0$, corresponds to the ring $x^2 + y^2 = a^2$. \\
One can analytically continue the Kerr solution for negative values of $r$ \cite{ch}. The horizons are at:
\b
r_\pm = \alpha \pm \sqrt{\alpha^2 - a^2} \, .
\e
The equations of the corresponding static horizons are:
\b
r_\pm(\theta) = \alpha \pm \sqrt{\alpha^2 - a^2 \cos^2 \theta} \, .
\e
There are no timelike coordinates inside the ergosphere --- the region between the event horizon and surrounding
static horizon. \\
Then Kerr solution describes two universes which behave asymptotically as Schwarzschild universes --- one with
$r > 0$ and having a positive centre $\alpha$, event horizon at $r_+$, and a Cauchy horizon at $r_- \, $; the
other --- with $r < 0$ and having a negative centre $\alpha$, event horizon at $r_-$, and a Cauchy horizon at
$r_+$. In our context this has the natural interpretation of a black hole solution with positive/negative charge
$q = \alpha \beta$  and a magnetic dipole of strength $m =a \alpha \beta$.

\section{Four-dimensional Lorentzian slice with Taub--NUT Geometry}

We consider five-dimensional Kaluza's universe with a four-dimensional \linebreak Lorentzian Taub-NUT \cite{Taub, NUT}
slice:
\b
d \sigma^2 & = & - V(r)(dt + 2 \ell \cos\theta \, d\phi)^2 + \frac{1}{V(r)} \, dr^2 \nonumber \\
& & \hskip30pt + \, (r^2 + \ell^2) (d \theta^2 + \sin^2 \theta \, d\phi^2) + ds^2 \, ,
\e
where
\b
V(r) = 1 - 2 \, \frac{\alpha r + \ell^2}{r^2 + \ell^2} \, ,
\e
where $\alpha$ and $\ell$ are, again, integration constants. \\
This metric has conical singularities at $\theta = 0, \, \pi$ (Misner string \cite{Misner}). The event horizon is
where $V(r)$ vanishes, i.e. at:
\b
r_\pm = \alpha \pm \sqrt{\alpha^2 + \ell^2} \, .
\e
The metric can be analytically continued for negative $r$ in a similar way and we will be interested in the two regions:
Region I with $\alpha \le 0$ and $r < r_- < 0$ and Region II with $\alpha \ge 0$ and $r > r_+ > 0$. \\
There is also an "ergoregion", surrounding the Misner string, inside of which $\phi$ is another timelike coordinate. The
equation of these horizons is:
\b
\tan^2 \theta = \frac{4 \ell^2 V(r)}{r^2 + \ell^2} \, ,  \: \: r < r_- \, \mbox{ or } \, r > r_+ \, .
\e
For the electromagnetic potentials (\ref{pot}) introduced with the transformation (\ref{shift}), we get:
\b
\label{hop1}
A_t & = & - \beta V(r) \, , \\
A_r & = & 0 \, , \\
A_\theta & = & 0 \, , \\
\label{hop2}
A_\phi & = & - 2 \ell \beta V(r) \cos \theta \, .
\e
Asymptotically, for large $r$ we have:
\b
\label{electric}
A_t & = & - \beta \left(1 - \frac{2 \alpha}{r} \right) \, , \\
\label{magnetic}
A_\phi & = & - 2 \ell \beta \left(1 - \frac{2 \alpha}{r} \right) \cos \theta \, .
\e
Equation (\ref{electric}) is the electric potential due to charge
\b
\label{qtaub}
q = 2 \alpha \beta \, .
\e
Equation (\ref{magnetic}) is the potential due to a magnetic monopole of charge $m = 2 \beta \ell$. This can be seen by
integrating the flux of the magnetic field $B^r$ through the infinite sphere:
\b
\label{mtaub}
4 \pi m = \lim_{r \to \infty} \oint_{S_r} B^r \, ds_r = \lim_{r \to \infty} \oint_{S_r} F_{\theta \phi} \, d\theta d\phi
= 4 \pi (2 \beta \ell) \, .
\e
For Kerr geometry this integral vanishes (we do not have a monopole but a magnetic dipole there). \\
The location of these charges appears to be unclear as $V(r) = 1 - 2 (\alpha r + \ell^2)(r^2 + \ell^2)^{-1}$, appearing
in (\ref{hop1}) and (\ref{hop2}), is not singular outside the horizons. However, every spacelike hypersurface which
is pushed between the horizons becomes singular \cite{Brill} and therefore, we can interpret the points
$\, ir = \pm \ell \, $ (note that $ir$ is a spacelike coordinate between the horizons) as the loci where the images of
the charges are seen by an observer with $r > r_+$ or $r < r_-$. For the case $\alpha = 0$, the proper distance
between the origin and the location of the images is:
\b
\int\limits_{i0}^{\pm i \ell} \frac{dr}{\sqrt{V(r)}} \,  =
\pm \, \, \ell \!\!\!\!\! \int\limits_{0}^{\ln (1 + \sqrt{2})} \sqrt{1 - \sinh ^2 x} \, dx \: \approx
\: \pm \,  0.7 \ell \, .
\e
Therefore, an observer with $r > r_+ > 0$ (Region I) will register the image of a monopole a proper distance
$\vert 0.7 \ell \vert$ from the origin, while an observer with $r < r_- < 0$ (Region II) will register the image of
a monopole a proper distance $-\vert 0.7 \ell \vert$ from the origin.

\section{Conclusions}
We have presented solutions to the original Kaluza's theory which describe static electric and magnetic fields
generated by point-like electric and magnetic charges and dipoles. Unlike the dual Kaluza--Klein theory (namely,
Klein's modification of the original Kaluza's theory) these solutions allow to have a constant Newton's constant,
as the gauge degrees of freedom are now transferred to the dilaton. The gauge-fixing of the electromagnetic
potentials results in the appearance of a Belinfante ghost part in the full energy-momentum tensor which fully
compensates the electromagnetic energy-momentum tensor. Four-dimensional solutions with vanishing Ricci tensor
($r_{ij} = 0$) are the only possible solutions when the dilaton is required to be constant in a Ricci-flat
five-dimensional universe. These four-dimensional gravitational Ricci-flat solutions can be interpreted from a
five-dimensional Kaluza's perspective, as solutions generated by four-dimensional electromagnetism of charges
and dipoles or their images (for the case of a Taub--NUT four-dimensional slice).  The integration constants
$a \, , \alpha  \mbox{ and } \beta$ (for Kerr geometry) and $\alpha \, , \ell  \mbox{ and } \beta$ (for Taub--NUT
geometry) have interpretation as charges (see (\ref{qkerr}) and (\ref{mkerr}) for Kerr geometry and
(\ref{qtaub}) and (\ref{mtaub}) for Taub--NUT geometry) and the solutions represent gravitational attraction
without unphysical regions with gravitational repulsion, as, for example, in the Reissner--Nordstr\o m
case \cite{MTW}.

\section*{Acknowledgements}
\noindent
We would like to thank Siddhartha Sen, Brian Nolan and Vesselin Gueorguiev for very helpful discussions.

\end{document}